\documentclass[aps,pra,preprint,groupedaddress,showpacs,floatfix]{revtex4}
\usepackage{dcolumn}
\usepackage{amsmath}
\usepackage{bm}
\usepackage{times}
\newcolumntype{.}{D{x}{}{-1}}

\newcommand*{\centt}[1]{\multicolumn{1}{c}{#1}}
\newcolumntype{w}[1]{D{.}{.}{#1}}

\begin{document}
\preprint{Version 1.3}

\title{Helium energy levels including $m \alpha^6$ corrections}

\author{Krzysztof Pachucki}
\email[]{krp@fuw.edu.pl} \homepage[]{www.fuw.edu.pl/~krp}

\affiliation{Institute of Theoretical Physics, Warsaw University,
             Ho\.{z}a 69, 00-681 Warsaw, Poland}

\date{\today}

\begin{abstract}
The $m\,\alpha^6$ correction to energy is expressed in terms of an effective
Hamiltonian $H^{(6)}$ for an arbitrary state of helium. Numerical
calculations are performed for $n=2$ levels, and the previous result
for the $2^3P$ centroid is corrected. While the resulting theoretical 
predictions for the ionization energy are in moderate agreement 
with experimental values for $2^3S_1$, $2^3P$, and $2^1S_0$ states, 
they are in significant disagreement for the singlet state $2^1P_1$.
\end{abstract}

\pacs{31.30.J-, 12.20.Ds, 31.15.xp, 32.10.Fn} \maketitle

High precision calculations of helium energy levels including relativistic and
QED effects is a complicated task \cite{hand}. It has been recognized early on
that the two-electron Dirac-Coulomb Hamiltonian is only an approximate
Hamiltonian, as it includes negative energy spectra and
does not account for magnetic and higher order interactions between
electrons. The proper approach has to be based on Quantum Electrodynamic theory.
For heavy few-electron ions the interactions between electrons can be treated 
perturbatively, on the same footing as the electron self-energy and 
vacuum polarization. Highly accurate results have 
been obtained for heavy helium- and lithium-like ions \cite{heavy1, heavy2},
and a convenient formulation of this  $1/Z$ expansion has been introduced
a few years ago by Shabaev in \cite{shabaev}. For systems with a larger
number of electrons the zeroth order Hamiltonian will include
an effective local potential to approximately account for interactions
between electrons. This approach is being developed by Sapirstein {\em et al} 
\cite{qed1}, and more recently by Shabaev and collaborators \cite{qed2}. 
One of the most interesting results obtained so far was the calculation 
of QED corrections to parity violation in the cesium atom \cite{qed2}.

For light atomic systems relativistic and QED effects are 
only a small correction  to the nonrelativistic Hamiltonian,
and for this reason they can be treated perturbatively.
More precisely, this perturbative approach relies on expansion of the binding energy 
in powers of the fine structure constant $\alpha$
\begin{equation}
E(\alpha) = E^{(2)} + E^{(4)} + E^{(5)} + E^{(6)} + E^{(7)} + O(\alpha^8),
\end{equation}
where $E^{(n)} = m\,\alpha^n\,{\cal E}^{(n)} $ is a contribution of order
$\alpha^n$. However, this expansion is nonanalytic, inasmuch as some of the
${\cal E}^{(n)}$ coefficients contain $\ln\alpha$, see for example Eq. (\ref{04}). 
Each ${\cal E}^{(n)}$ can be expressed in terms of 
the expectation value of some effective Hamiltonian $H^{(n)}$ with 
the nonrelativistic wave function \cite{fw}. This approach allows for a consistent 
inclusion of all relativistic and QED effects order by order in $\alpha$. 
We present in this work high precision calculations of $n=2$ energy levels in
helium including the contribution $E^{(6)}$. This contribution
has already been derived separately for triplet states in \cite{dk,numstrip}, 
and for singlet states in \cite{ky,kp}. 
Here we obtain $H^{(6)}$ valid for all helium states,
and present numerical results for $2^3S_1, 2^3P, 2^1S_0$ and $2^1P_1$ energy levels.

The leading term in the expansion of the energy in powers of $\alpha$, 
${\cal E}^{(2)} = {\cal E}$, is the nonrelativistic energy, 
the eigenvalue of the nonrelativistic  Hamiltonian, which in atomic units is
\begin{equation}
H^{(2)} \equiv H = \sum_a\biggl[\frac{\vec p_a^{\;2}}{2}-\frac{Z}{r_a}\biggr] +
\sum_{a>b}\frac{1}{r_{ab}}
\end{equation}
The relativistic correction ${\cal E}^{(4)}$ is the expectation value
of the Breit-Pauli Hamiltonian $H^{(4)}$ \cite{bethe}
\begin{eqnarray}
H^{(4)} &=&\sum_a \biggl\{-\frac{\vec p^{\;4}_a}{8} +
\frac{ \pi\,Z}{2}\,\delta^3(r_a)
+\frac{Z}{4}\,
\vec\sigma_a\cdot\frac{\vec r_a}{r_a^3}\times \vec p_a\biggr\}
\nonumber \\
&& +\sum_{a>b} \biggl\{
\pi\, \delta^3(r_{ab})
-\frac{1}{2}\, p_a^i\,
\biggl(\frac{\delta^{ij}}{r_{ab}}+\frac{r^i_{ab}\,r^j_{ab}}{r^3_{ab}}
\biggr)\, p_b^j 
+\frac{\sigma_a^i\,\sigma_b^j}{4\,r_{ab}^3}\,
\biggl(\delta^{ij}-3\,\frac{r_{ab}^i\,r_{ab}^j}{r_{ab}^2}\biggr)
\nonumber \\ && 
+\frac{1}{4\,r_{ab}^3} \biggl[
2\,\bigl(\vec\sigma_a\cdot\vec r_{ab}\times\vec p_b -
\vec\sigma_b\cdot\vec r_{ab}\times\vec p_a\bigr)+
\bigl(\vec\sigma_b\cdot\vec r_{ab}\times\vec p_b -
\vec\sigma_a\cdot\vec
r_{ab}\times\vec p_a\bigr)\biggr]\biggr\}\,. \label{03}
\end{eqnarray}

${\cal E}^{(5)}$ is the leading QED correction. Apart from the anomalous magnetic
moment correction to the spin-orbit and spin-spin interactions,
which we neglect here, as we consider singlet or spin-orbit averaged (centroid)
levels, it includes the following terms \cite{hand}
\begin{eqnarray}
{\cal E}^{(5)} &=&\sum_{a>b}\Biggl\langle
\left[\frac{164}{15}+\frac{14}{3}\,\ln\alpha\right]\,\delta^3(r_{ab})
-\frac{7}{6\,\pi}\,\frac{1}{r_{ab}^3}\Biggr\rangle \label{04}\\
&& +\sum_a \left[\frac{19}{30}+\ln(\alpha^{-2})-\ln k_0\right]
   \frac{4\,Z}{3} \langle{\delta^3(r_a)}\rangle, \nonumber
\end{eqnarray}
where
\begin{eqnarray} 
\left\langle\frac{1}{r^3}\right\rangle &\equiv&
\lim_{a\rightarrow 0}\int d^3 r\, \phi^{*}(\vec r)\,\phi(\vec r)
\left[\frac{1}{r^3}\,\Theta(r-a) + 4\,\pi\,\delta^3(r)\,
(\gamma+\ln a)\right],\\
\ln k_0 & = & \frac{\Bigl\langle\sum_a \vec p_a\,(H-{\cal E})\,
  \ln\bigl[2\,(H-{\cal E})\bigr]\,
  \sum_b\vec p_b\Bigr\rangle}{2\,\pi\,Z\,
  \Bigl\langle\sum_c\delta^3(r_c)\Bigr\rangle}.
\end{eqnarray} 
The next order contribution ${\cal E}^{(6)}$ is much more complicated. 
It can be represented in general as
\begin{equation}
{\cal E}^{(6)} = \langle H^{(6)}\rangle + \biggl\langle H^{(4)}\,
\frac{1}{({\cal E}-H)'}\, H^{(4)}\biggr\rangle \label{07}
\end{equation}
but separate matrix elements of the first and the second
term in the above are divergent. The spin dependent terms which contribute 
to fine structure are finite, and have been derived by
Douglas and Kroll in \cite{dk}. These contributions are not included here, because
we consider spin-orbit averaged levels. The singularities of matrix elements
in Eq. (\ref{07}) can be eliminated by algebraic transformations \cite{kp}
in a similar way for both singlet and triplet states.
Therefore we extend the result obtained in \cite{kp} to arbitrary
states of helium, and the contribution ${\cal E}^{(6)}$ can be represented as
\begin{eqnarray}
{\cal E}^{(6)} &=& \biggl\langle -\frac{{\cal E}^3}{2} 
+\biggl[\biggl(-{\cal E} + \frac{3}{2}\,\vec p_2^{\;2} 
+\frac{1-2\,Z}{r_2}\biggr)\,\frac{Z\,\pi}{4}\,\delta^{3}(r_1)
+(1\leftrightarrow 2)\biggr]
\nonumber \\ &&
+\frac{\vec P^2}{6}\,\pi\,\delta^{3}(r) 
-\frac{(3+\vec\sigma_1\cdot\vec\sigma_2)}{24}\,\pi\,\vec p\,\delta^3(r)\,\vec p
-\biggl(\frac{Z}{r_1} + \frac{Z}{r_2}\biggr)\,\frac{\pi}{2}\,\delta^{3}(r)
\nonumber \\ &&
+\biggl(\frac{13}{12}+\frac{8}{\pi^2}-\frac{3}{2}\,\ln(2)
-\frac{39\,\zeta(3)}{4\,\pi^2}\biggr)\,\pi\,\delta^{3}(r)
+\frac{{\cal E}^2+2\,{\cal E}^{(4)}}{4\,r} 
\nonumber \\ &&
-\frac{{\cal E}}{r^2}\,\frac{(31+5\,\vec\sigma_1\cdot\vec\sigma_2)}{32} 
-\frac{{\cal E}}{2\,r}\,\bigg(\frac{Z}{r_1}+\frac{Z}{r_2}\biggr)
+\frac{{\cal E}}{4}\,\biggl(\frac{Z}{r_1}+\frac{Z}{r_2}\biggr)^2
\nonumber \\ &&
-\frac{1}{r^2}\,\biggl(\frac{Z}{r_1}+\frac{Z}{r_2}-\frac{1}{r}\biggr)\,
\frac{(23 + 5\,\vec\sigma_1\cdot\vec\sigma_2)}{32}
-\frac{1}{4\,r}\,\biggl(\frac{Z}{r_1}+\frac{Z}{r_2}\biggr)^2
\nonumber \\ &&
+\frac{Z^2}{2\,r_1\,r_2}\,\biggl({\cal E}+\frac{Z}{r_1}+\frac{Z}{r_2}-\frac{1}{r}\biggr)
-Z\,\biggl(\frac{\vec r_1}{r_1^3} - \frac{\vec
  r_2}{r_2^3}\biggr)\cdot\frac{\vec r}{r^3}\,
\frac{(13+5\,\vec\sigma_1\cdot\vec\sigma_2)}{64}
\nonumber \\ &&
+\frac{Z}{4}\,\biggl(\frac{\vec r_1}{r_1^3} - \frac{\vec r_2}{r_2^3}\biggr)\cdot\frac{\vec r}{r^2}
-\frac{Z^2}{8}\,\frac{r_1^i}{r_1^3}\,\frac{(r^i r^j - 3\,\delta^{ij}\,r^2)}{r}\,\frac{r_2^j}{r_2^3}
\nonumber \\ &&
+\biggl[\frac{Z^2}{8}\,\frac{1}{r_1^2}\,\vec p_2^{\;2} + 
\frac{Z^2}{8}\,\vec p_1\,\frac{1}{r_1^2}\,\vec p_1
+\vec p_1\,\frac{1}{r^2}\,\vec p_1\,\frac{(47 + 5\,\vec\sigma_1\cdot\vec\sigma_2)}{64}
+(1\leftrightarrow 2)\biggr]
\nonumber \\ &&
+\frac{1}{4}\,p_1^i\,\biggl(\frac{Z}{r_1}+\frac{Z}{r_2}\biggr)\,
\frac{(r^i\,r^j + \delta^{ij}\, r^2)}{r^3}\, p_2^j
+P^i\,\frac{(3\,r^i\,r^j - \delta^{ij} r^2)}{r^5}\,P^j\,\frac{(-3+\vec\sigma_1\cdot\vec\sigma_2)}{192}
\nonumber \\ &&
-\biggl[\frac{Z}{8}\,p_2^k\,\frac{r_1^i}{r_1^3}\,\biggl(\delta^{jk}\,\frac{r^i}{r} 
- \delta^{ik}\, \frac{r^j}{r} - \delta^{ij}\, \frac{r^k}{r} 
- \frac{r^i\, r^j\, r^k}{r^3}\biggr)\,p_2^j
+(1\leftrightarrow 2)\biggr]
\nonumber \\ &&
-\frac{{\cal E}}{8}\,p_1^2\,p_2^2-\frac{1}{4}\,p_1^2\,\biggl(\frac{Z}{r_1}
+\frac{Z}{r_2}\biggr)\,p_2^2
+\frac{1}{4}\,\vec p_1\times\vec p_2\,\frac{1}{r}\,\vec p_1\times\vec p_2
\nonumber \\ &&
+\frac{1}{8}\,p_1^k\,p_2^l\,\biggl(-\delta^{jl}\,\frac{r^i\,r^k}{r^3} -
           \delta^{ik}\,\frac{r^j\,r^l}{r^3} +
           3\,\frac{r^i\,r^j\,r^k\,r^l}{r^5} \biggr)\, p_1^i\,p_2^j\biggr\rangle
\nonumber \\ &&
+ E_{\rm sec} + E_{R1} + E_{R2} -\ln(\alpha)\,\pi\,\langle\delta^3(r)\rangle,\label{08}
\end{eqnarray}
where $\vec P = \vec p_1 + \vec p_2$, $\vec p = (\vec p_1-\vec p_2)/2$, $\vec
r = \vec r_1-\vec r_2$, and 
\begin{eqnarray}
E_{\rm sec} &=& \biggl\langle H'_A\,\frac{1}{({\cal E}-H)'}\,H'_A\biggr\rangle
 + \biggl\langle H_B\,\frac{1}{({\cal E}-H)'}\,H_B\biggr\rangle 
\nonumber \\ &&
+\biggl\langle H_C\,\frac{1}{{\cal E}-H}\,H_C\biggr\rangle 
+\biggl\langle H_D\,\frac{1}{({\cal E}-H)'}\,H_D\biggr\rangle
\end{eqnarray}
The operators $H'_A$, $H_B$, $H_C$ and $H_D$ are parts of the $H^{(4)}$
Hamiltonian from Eq. (\ref{03}), which was transformed \cite{kp}
to eliminate singularities from second order matrix elements 
\begin{eqnarray}
H'_A & = & -\frac{1}{2}\,({\cal E}-V)^2
-p_1^i\,\frac{1}{2\,r}\,\biggl(\delta^{ij}+\frac{r^i\,r^j}{r^2}\biggr)\,p_2^j
+\frac{1}{4}\,\vec \nabla_1^2 \, \vec \nabla_2^2
-\frac{Z}{4}\,\frac{\vec r_1}{r_1^3}\cdot\vec \nabla_1
-\frac{Z}{4}\,\frac{\vec r_1}{r_1^3}\cdot\vec \nabla_1 \label{11}\\ \nonumber \\
H_B & = & \left[
\frac{Z}{4}\biggl(
\frac{{\vec  r}_1}{r_1^3}\times{\vec p}_1+
\frac{{\vec r}_2}{r_2^3}\times{\vec p}_2\biggr)-
\frac{3}{4}\,\frac{{\vec r}}{r^3}\times
({\vec p}_1-{\vec p}_2)\right]\,\frac{{\vec \sigma}_1+{\vec \sigma}_2}{2}\,,\\
\nonumber \\
H_C & = & \left[
\frac{Z}{4}\biggl(
\frac{{\vec r}_1}{r_1^3}\times{\vec p}_1-
\frac{{\vec r}_2}{r_2^3}\times{\vec p}_2\biggr)+
\frac{1}{4}\,\frac{{\vec r}}{r^3}\times
({\vec p}_1+{\vec p}_2)\right]\,\frac{{\vec \sigma}_1-{\vec \sigma}_2}{2}\,,\\
\nonumber \\
H_D  & = & \frac{1}{4}\left(
\frac{{\vec \sigma}_1\,{\vec \sigma}_2}{r^3}
-3\,\frac{{\vec \sigma}_1\cdot{\vec r}\;
{\vec \sigma}_2\cdot{\vec r}}{r^5}\right)\,.
\end{eqnarray}
where $\vec \nabla_1^2 \, \vec \nabla_2^2$ in $H'_A$ in Eq. (\ref{11}) 
is understood as a differentiation
of $\phi$ on the right hand side as a function (omitting $\delta^3(r)$).
$E_{R1}$ and $E_{R2}$ are one- and two-loop electron self-energy and vacuum
polarization corrections respectively \cite{ky,kp},
\begin{eqnarray}
E_{R1} &=& Z^2\,\biggl[\frac{427}{96}-2\,\ln(2)\biggr]\,\pi\,
\langle\delta^3(r_1)+\delta^3(r_2)\rangle
\nonumber \\ &&
+\biggl[
\frac{6\,\zeta(3)}{\pi^2}-\frac{697}{27\,\pi^2}-8\,\ln(2)+\frac{1099}{72}
\biggr]\,\pi\,\langle\delta^3(r)\rangle, \\ \nonumber \\
E_{R2} &=& Z\,\biggl[-\frac{9\,\zeta(3)}{4\,\pi^2}-\frac{2179}{648\,\pi^2}+
\frac{3\,\ln(2)}{2}-\frac{10}{27}\biggr]\,\pi\,\langle\delta^3(r_1)+\delta^3(r_2)\rangle
\nonumber \\ &&
+\biggl[\frac{15\,\zeta(3)}{2\,\pi^2}+\frac{631}{54\,\pi^2}-5\,\ln(2)+\frac{29}{27}
\biggr]\,\pi\,\langle\delta^3(r)\rangle, 
\end{eqnarray}

The higher order contribution ${\cal E}^{(7)}$ is known only to some approximation. Following
Ref. \cite{md} the hydrogenic values for one-, two-, and three-loop contributions \cite{eides}
at order $m\,\alpha^7$ are extrapolated to helium, according to 
\begin{equation}
{\cal E}^{(7)} = \bigl[{\cal E}^{(7)}(1S,{\rm He}^+) +{\cal E}^{(7)}(nX,{\rm He}^+)\bigr]  
\,\frac{\langle \delta^3(r_1) + \delta^3(r_2)\rangle_{\rm He}}
{\langle \delta^3(r)\rangle_{1S,{\rm He}^+} 
+ \langle\delta^3(r)\rangle_{nX,{\rm He}^+}} - {\cal E}^{(7)}(1S,{\rm He}^+)
\label{17}
\end{equation}
for $X=S$, and for states with higher angular momenta 
${\cal E}^{(7)}(nX,{\rm He}^+)$ is neglected. 

We pass now to the calculation of matrix elements. The wave function
is expressed in terms of explicitly correlated exponential functions $\phi_i$ 
\begin{eqnarray}
\phi_i &=& e^{-\alpha_i\,r_1 - \beta_i\,r_2 - \gamma_i\,r_{12}} \pm
(r_1\leftrightarrow r_2)\\
\vec \phi_i &=& \vec r_1\,e^{-\alpha_i\,r_1 - \beta_i\,r_2 - \gamma_i\,r_{12}} \pm
(r_1\leftrightarrow r_2)
\end{eqnarray}
with random $\alpha_i,\beta_i,\gamma_i$ \cite{kor1}. This basis set is a very
effective representation of the two-electron wave function, so much so that
the nonrelativistic energies with $1500$ basis functions are accurate to about 18 digits.
Moreover, matrix elements of operators for relativistic and higher order
corrections can all be obtained analytically in terms of rational,
logarithmic and dilogarithmic functions, for example
\begin{equation}
\frac{1}{16\,\pi^2}\,\int d^3 r_1\,\int d^3r_2\,
\frac{e^{-\alpha r_1-\beta r_2-\gamma r}}{r_1\,r_2\,r} =
\frac{1}{(\alpha+\beta)(\beta+\gamma)(\gamma+\alpha)}.
\end{equation}
Numerical results 
for matrix elements of $m\,\alpha^6$
operators with singlet and triplet $P$ states are presented in Table I:
due to the singularity of these operators we had to use octuple precision arithmetic.
\begin{table}[htb]
\renewcommand{\arraystretch}{0.90}
\caption{Expectation values of operators entering $H^{(6)}$ for the
  $2^1P_1$ and $2^3P$ centroid.}
\label{TBL1}
\begin{tabular}{l@{\hspace{0.2cm}}.@{\hspace{1.0cm}}.}
\hline
\hline
\multicolumn{1}{l}{operator} &
\multicolumn{1}{c}{$2^1P_1$} &
\multicolumn{1}{c}{$2^3P$} \\
\hline
$ 4\,\pi\,\delta^{3}(r_1)         $& 16x.014\,493 & 15x.819\,309 \\ 
$ 4\,\pi\,\delta^{3}(r)           $&  0x.009\,238 &  0x.0       \\ 
$ 4\,\pi\,\delta^{3}(r_1)/r_2     $&  3x.934\,080 &  4x.349\,766 \\ 
$ 4\,\pi\,\delta^{3}(r_1)\,p_2^2  $&  3x.866\,237 &  4x.792\,830 \\ 
$ 4\,\pi\,\delta^{3}(r)/r_1       $&  0x.012\,785 &  0x.0       \\ 
$ 4\,\pi\,\delta^{3}(r)\,P^2      $&  0x.070\,787 &  0x.0       \\ 
$ 4\,\pi\,\vec p\,\delta^{3}(r)\,\vec p $
                                   &  0x.0       &  0x.077\,524 \\ 
$ 1/r                             $&  0x.245\,024 &  0x.266\,641 \\ 
$ 1/r^2                           $&  0x.085\,798 &  0x.094\,057 \\ 
$ 1/r^3                           $&  0x.042\,405 &  0x.047\,927 \\ 
$ 1/r_1^2                         $&  4x.043\,035 &  4x.014\,865 \\ 
$ 1/(r_1\,r_2)                    $&  0x.491\,245 &  0x.550\,342 \\ 
$ 1/(r_1\,r)                      $&  0x.285\,360 &  0x.317\,639 \\ 
$ 1/(r_1\,r_2\,r)                 $&  0x.159\,885 &  0x.198\,346 \\ 
$ 1/(r_1^2\,r_2)                  $&  1x.063\,079 &  1x.196\,631 \\ 
$ 1/(r_1^2\,r)                    $&  1x.002\,157 &  1x.109\,463 \\ 
$ 1/(r_1\,r^2)                    $&  0x.105\,081 &  0x.121\,112 \\ 
$ (\vec r_1\cdot \vec r)/(r_1^3\,r^3) 
                                  $&  0x.010\,472 &  0x.030\,284 \\ 
$ (\vec r_1\cdot \vec r)/(r_1^3\,r^2) 
                                  $&  0x.043\,524 &  0x.075\,373 \\ 
$ r_1^i\,r_2^j\,(r^i r^j - 3\,\delta^{ij}\,r^2)/(r_1^3\,r_2^3\,r)  
                                  $& -0x.004\,745 &  0x.090\,381 \\ 
$ p_2^2/r_1^2                     $&  1x.127\,058 &  1x.410\,228 \\ 
$ \vec p_1\, /r_1^2\, \vec p_1    $& 16x.067\,214 & 15x.925\,672 \\ 
$ \vec p_1\, /r^2\, \vec p_1      $&  0x.190\,797 &  0x.279\,229 \\ 
$ p_1^i\,(r^i\,r^j + \delta^{ij}\, r^2)/(r_1\,r^3)\, p_2^j            
                                  $&  0x.053\,432 & -0x.097\,364 \\ 
$ P^i\,(3\,r^i\,r^j - \delta^{ij} r^2)/r^5\,P^j                           
                                  $&  0x.013\,743 & -0x.060\,473 \\ 
$  p_2^k\, r_1^i\,/r_1^3\,(\delta^{jk}\, r^i/r - \delta^{ik}\, r^j/r - 
           \delta^{ij}\, r^k/r - r^i\, r^j\, r^k/r^3)\,p_2^j
                                  $& -0x.039\,975 &  0x.071\,600 \\ 
$ p_1^2\,p_2^2                    $&  0x.973\,055 &  1x.198\,492 \\ 
$ p_1^2\,/r_1\,p_2^2              $&  3x.102\,248 &  3x.883\,404 \\ 
$ \vec p_1\times\vec p_2\,/r\,\vec p_1\times\vec p_2           
                                  $&  0x.216\,869 &  0x.399\,306 \\ 
$ p_1^k\,p_2^l\,(-\delta^{jl}\,r^i\,r^k/r^3 -
           \delta^{ik}\,r^j\,r^l/r^3 + 3\,r^i\,r^j\,r^k\,r^l/r^5)\, p_1^i\,p_2^j   
                                  $& -0x.126\,416 & -0x.187\,304 \\ 
\hline
\hline
\end{tabular}
\end{table}
The $m\,\alpha^6$ correction to the energy also involves second order matrix
elements $E_{\rm sec}$. If we write
\begin{eqnarray}
H'_A &=&      Q_A \\
H_B &=& \vec Q_B\cdot\vec s\\
H_C &=& \vec Q_C\cdot\frac{(\vec\sigma_1-\vec\sigma_2)}{2}\\
H_D &=& Q_D^{ij}\,s^i\,s^j
\end{eqnarray}
then one obtains for singlet states 
\begin{eqnarray}
E(2^1S_0)_{\rm sec} &=& \langle 2^1S|Q_A\,\frac{1}{{\cal E}-H}\,Q_A |2^1S\rangle+
              \langle 2^1S|Q_C^j\,\frac{1}{{\cal E}-H}\,Q_C^j |2^1S\rangle\\
E(2^1P_1)_{\rm sec} &=& \langle 2^1P^i| Q_A\,\frac{1}{{\cal E}-H}\,Q_A |2^1P^i\rangle+
               \langle 2^1P^i| Q_C^j\,\frac{1}{{\cal E}-H}\,Q_C^j |2^1P^i\rangle
\end{eqnarray}
and the contributions from $H_B$ and $H_D$ vanish.
The result for the $2^3S_1$ state is
\begin{eqnarray}
E(2^3S_1)_{\rm sec} &=& \langle 2^3S|Q_A\,\frac{1}{{\cal E}-H}\,Q_A |2^3S\rangle 
              +\frac{2}{3}\,\langle 2^3S|Q_B^j\,\frac{1}{{\cal E}-H}\,Q_B^j
              |2^3S\rangle \nonumber \\ 
          &+& \frac{1}{3}\,\langle 2^3S|Q_C^j\,\frac{1}{{\cal E}-H}\,Q_C^j |2^3S\rangle
              +\frac{1}{3}\,\langle 2^3S|Q_D^{ij}\,\frac{1}{{\cal E}-H}\,Q_D^{ij} |2^3S\rangle
\end{eqnarray}
The result for the $2^3P$ centroid, defined by
\begin{equation}
E(2^3P) = \frac{1}{9}\,\left[E(2^3P_0) + 3\,E(2^3P_1) + 5\,E(2^3P_2)\right]
\end{equation}
is
\begin{eqnarray}
E(2^3P)_{\rm sec} &=& \langle 2^3P^i| Q_A\,\frac{1}{{\cal E}-H}\,Q_A |2^3P^i\rangle+
                     \frac{2}{3}\,\langle 2^3P^i| Q_B^j\,\frac{1}{{\cal E}-H}\,Q_B^j
                     |2^3P^i\rangle \nonumber\\
                 &+& \frac{1}{3}\,\langle 2^3P^i| Q_C^j\,\frac{1}{{\cal E}-H}\,Q_C^j |2^3P^i\rangle+
                 \frac{1}{3}\,\langle 2^3P^i| Q_D^{jk}\,\frac{1}{{\cal E}-H}\,Q_D^{jk} |2^3P^i\rangle
\end{eqnarray}
Numerical results for second order matrix elements are presented in Table II. 
\begin{table}[!htb]
\renewcommand{\arraystretch}{0.8}
\caption{Contributions to ionization energy ${\cal E}^{(6)}$ 
         for $n=2$ states of the helium atom. $E_Q$ is a sum of operators in
         Eq. (\ref{08}), in comparison to Ref. \cite{kp} it includes
         the contribution $E_H$. $E_{LG}$ is the logarithmic contribution,
         last term in Eq. (\ref{08}). The sum of spin dependent, second order 
         corrections for $2^3P$ centroid is taken from Ref. \cite{plamb}.}
\label{TBL2}
\begin{ruledtabular}
\begin{tabular}{c@{\extracolsep{\fill}}w{3.11}w{3.11}w{3.8}w{3.11}}
$m\,\alpha^6$         & \centt{$2^1$S}&\centt{$2^1$P}&\centt{$2^3$S}&\centt{$2^3$P} \\
\hline
$E_Q$                 &  12.287\,491      &  12.236\,966      &  13.052\,109     &  11.963\,305     \\
$E'_A$                & -16.280\,186(10)  & -16.084\,034(5)   & -17.189\,809(10) & -15.848\,510(2)  \\
$E_B$                 &   0.0             &   0.0             &  -0.018\,722     &            \\[-1.5ex]
$E_C$                 &  -0.033\,790      &   0.201\,363      &  -0.001\,108     & \Biggr\} -0.168\,704(2)  \\[-1.5ex]
$E_D$                 &   0.0             &   0.0             &  -0.003\,848     &           \\
$-E_{\rm Dirac}({\rm He}^+)$
                      &   4.000\,000      &   4.000\,000      &   4.000\,000     &   4.000\,000    \\ 
\\
Subtotal              &  -0.026\,485(10)  &   0.354\,296(5)   &  -0.161\,377(10) &  -0.053\,908(3)  \\ 
$E_{R1}$              &   2.999\,960      &   0.106\,839      &   3.625\,397     &  -1.106\,416     \\
$E_{R2}$              &   0.016\,860      &   0.000\,112      &   0.032\,331     &  -0.009\,867     \\
$E_{LG}$              &   0.133\,682      &   0.011\,364      &   0.0            &   0.0           \\ 
\\
Total                 &   3.124\,017(10)  &   0.472\,611(5)   &   3.496\,351(10) &  -1.170\,191(3)  \\ 
\end{tabular}
\end{ruledtabular}
\end{table}
One notices a strong cancellation between $m\,\alpha^6$
contributions and the Dirac energy for the He$^+$ ion, the subtotal line in Table
II. Because of this cancellation, the dominant contribution is the one loop
radiative correction, with the exception of the $2^1P_1$ state, where 
the wave function at the nucleus happens to be very close to 16, the He$^+$ value,
see Table I. 

The summary of all important contributions to ionization energies
is presented in Table III. 
\begin{table}[!htb]
\renewcommand{\arraystretch}{0.8}
\caption{Contributions to ionization energy of $n=2$ helium states in MHz. 
         Physical constants from \cite{nist}, $R_\infty = 10\,973\,731.568\,525(73)$ m$^{-1}$, 
         $\alpha  = 1/137.03599911(46)$, $\not\!\!\lambda_e = 386.1592678(26)$ fm,
         $m_\alpha/m_e = 7294.2995363(32)$,  $c = 299792458$. $E_{\rm fs}$ is
         a finite nuclear size correction with the charge radius $r_\alpha = 1.673$ fm.}
\label{TBL3}
\begin{ruledtabular}
\begin{tabular}{c@{\extracolsep{\fill}}w{10.7}w{11.7}w{11.8}w{10.5}w{12.5}}
                      & \centt{$\nu(2^1S)$} &  \centt{$\nu(2^1P)$} &
                      \centt{$\nu(2^3S)$} &  \centt{$\nu(2^3P)$}\\
\hline
 $\mu\,\alpha^2$      &   -960\,331\,428.61&   -814\,736\,669.94&   -1\,152\,795\,881.77&   -876\,058\,183.13 \\
 $\mu^2/M\,\alpha^2$  &         8\,570.43&        41\,522.20&          6\,711.19&       -58\,230.36 \\
 $\mu^3/M^2\,\alpha^2$&          -16.72&          -20.80&            -7.11&          -25.33 \\
 $E_{\rm fs}$         &            1.99&            0.06&             2.59&           -0.79 \\
 $m\,\alpha^4$        &       -11\,971.45&       -14\,024.05&        -57\,629.31&        11\,436.88 \\
 $m^2/M\,\alpha^4$    &           -3.34&           -2.81&             4.28&           11.05 \\
 $m\,\alpha^5$        &         2\,755.76&           38.77&          3\,999.43&        -1\,234.73 \\
 $m^2/M\,\alpha^5$    &           -0.63&            0.47&            -0.80&           -0.62 \\
 $m\,\alpha^6$        &           58.29&            8.82&            65.24&          -21.83 \\
 $m\,\alpha^7$        &         -3.85(1.90)&      -0.16(16)&         -5.31(1.00)&            1.93(40) \\
\\
 $E_{\rm the}$        & -960\,332\,038.13(1.90)& -814\,709\,147.44(16)&   -1\,152\,842\,741.56(1.00)&   -876\,106\,246.93(40)\\
 $E_{\rm exp}$        & -960\,332\,040.86(15)  & -814\,709\,153.0(3.0)&   -1\,152\,842\,742.97(0.06)&   -876\,106\,247.35(6)\\
\end{tabular}
\end{ruledtabular}
\end{table}
We include the first and second order
mass polarization correction to the nonrelativistic energy, as well as
first order nuclear recoil corrections $\alpha^4\,m^2/M$ and $\alpha^5\,m^2/M$.
We expect higher order terms in the mass ratio to be much below the $0.01$ MHz level,
the precision of calculated contributions, see Table III.
Results for nonrelativistic as well as for leading relativistic
corrections are in agreement with those obtained previously by Drake \cite{md, morton}. 
Corrections of order $m\,\alpha^5$ were calculated using the Drake and Goldman \cite{bethelog} 
values for Bethe logarithms. The $m\,\alpha^6$ correction is calculated in this work.
All but $m\,\alpha^7$ contributions are calculated exactly. This last one,
$m\,\alpha^7$ is estimated on the basis of the hydrogenic value according to
Eq. (\ref{17}). It is the only source of uncertainty of theoretical predictions,
as the achieved numerical precision for each correction is below $0.01$ MHz.

The value for the $2^1S_0$ state has already been presented in our former work \cite{kp};
here we display in more detail all the contributions. The value for the
$2^3S_1$ state is in agreement with our previous calculation in \cite{numstrip}, 
where we obtained ${\cal E}^{(6)} = 3.496\,93(50)$.
This provides justification of the correctness of the obtained result, since the two
derivations of the $m\,\alpha^6$ operators were performed in a different way.
However, the result for the $2^3P$ state is in disagreement with our result
from \cite{plamb}. For this reason we checked Ref. \cite{plamb}, and found a mistake.
The derived set of operators representing ${\cal E}^{(6)}$ was correct,
but the expectation value of $H'_{EN}$, in the notation of \cite{plamb}, was in error. 
The correct result is $\langle H'_{EN}\rangle = 11.903\,751$. With the second order
matrix element $-15.838\,656(9)$ and subtracting He$^+$ $m\,\alpha^6$ energy $-Z^6/16$,
it is equal to $0.049\,702(9)$, while the former result was $0.140\,689(9)$,
see Table II of \cite{plamb}. Together with other corrections from that Table
the total $m\,\alpha^6$ contribution becomes $-1.170\,188(9)$, in agreement
with the result obtained here.   

We find a moderate agreement with experimental ionization energies
for the $2^1S_0$, $2^3S_1$ and $2^3P$ states but 
a significant disagreement for the $2^1P_{1}$ state. 
Following \cite{md}, the result for the $2^3S_1$ state was obtained by combining the
$2^3S_1 - 3^3D_1$ measurement by Dorrer {\em et al} \cite{Dorrer}
$786\,823\,850.002(56)$ MHz with the theoretical $3^3D_1$ ionization energy
$366\,018\,892.97(2)$ MHz calculated by Drake \cite{md, morton}.
The ionization energy of the $2^3P$ state was obtained from the measurement 
of the $2^3S_1 - 2^3P$ transition by Cancio {\em et al} \cite{Cancio}
of $276\,736\,495.6246(24)$ MHz and the previously obtained $2^3S_1$ 
ionization energy. The ionization energy of the $2^1S$ state was obtained   
from measurements of $2^1S - n^1D$ transitions by Lichten {\em et al} \cite{Lichten}
with $n=7-20$ and Drake's calculations for $n^1D$ states \cite{md, morton}.
Finally, the result for $2^1P$ ionization energy is determined
by combining the $2^1P - 3^1D_2$ transition $448\,791\,404.0(30)$ MHz
by Sansonetti and Martin \cite{sans} (including correction of $0.6$ MHz \cite{md}),
with calculated \cite{md, morton} $3^1D_2$ energy $365\,917\,749.02(2)$ MHz.
The significant disagreement with theoretical predictions for $2^1P$
state calls for an independent calculation of the $m\,\alpha^6$ term, and 
on the other hand for the direct frequency measurement of $2^1P - 3^1D_2$ 
or $2^1P - 2^1S$ transitions.

Further improvement of theoretical predictions can be achieved by 
the calculation of $m\,\alpha^7$ contributions. The principal problem here
will be the numerical evaluation of the relativistic corrections 
to Bethe-logarithms and the derivation of remaining operators.
Such a calculation has recently been performed for helium 
fine structure \cite{fs}, therefore in view of 
newly proposed experiments \cite{vassen}, calculations 
for other states of helium although not simple, can be achieved.  

\section*{Acknowledgments}
The author wishes to acknowledge interesting discussions with Vladimir Korobov.
This work was supported in part by  Postdoctoral Training Program HPRN-CT-2002-0277.


\begin{thebibliography}{99}
\bibitem{hand} {\em Handbook of Atomic, Molecular and Optical Physics},
               Ed. G.~W.~F. Drake, Springer (2006).
\bibitem{heavy1} J. Sapirstein and K.T. Cheng, Phys. Rev. A {\bf 64}, 022502 (2001).
\bibitem{heavy2} V.A. Yerokhin, P. Indelicato and V. Shabaev, {\em submitted to} Can. J. Phys.
\bibitem{shabaev} V.~M.~Shabaev, Phys. Rep. {\bf 356}, 119 (2002).
\bibitem{qed1} J. Sapirstein and K. T. Cheng, Phys. Rev. A {\bf 66}, 042501
               (2002), ibid. {\bf 67}, 022512 (2003).
\bibitem{qed2} V.M. Shabaev, K. Pachucki, I.I. Tupitsyn and V.A. Yerokhin,
                 Phys. Rev. Lett. {\bf 94}, 213002 (2005).
\bibitem{fw} K. Pachucki, Phys. Rev. A  {\bf 71}, 012503 (2005).
\bibitem{dk} M. Douglas and N.M. Kroll, Ann. Phys. (N.Y) {\bf 82}, 89 (1974).
\bibitem{numstrip}K. Pachucki, Phys. Rev. Lett. {\bf 84}, 4561 (2000).
\bibitem{ky} V. Korobov and A. Yelkhovsky, Phys. Rev. Lett. {\bf 87}, 193003 (2001),
             A. Yelkhovsky, Phys. Rev. A {\bf 64}, 062104 (2001).
\bibitem{kp} K. Pachucki, Phys. Rev. A. {\bf 74}, 022512 (2006).
\bibitem{bethe} H.A. Bethe and E.E. Salpeter,
                {\em Quantum Mechanics Of One- And Two-Electron Atoms},\\
                Plenum Publishing Corporation, New York (1977).
\bibitem{md} G.W.F. Drake and W.C. Martin, Can. J. Phys. {\bf 76}, 679 (1998).
\bibitem{eides} M.I. Eides, H. Grotch, and V.A. Shelyuto,
                Phys. Rep. {\bf 342}, 63 (2001).
\bibitem{kor1} V.I. Korobov, Phys. Rev. A {\bf 61}, 064503 (2000), 
               Phys. Rev. A {\bf 66}, 024501 (2002).
\bibitem{morton} D.C. Morton, Q. Wu, and G.~W.~F~Drake, Can. J. Phys. {\bf 82}, 835 (2005).
\bibitem{bethelog} G.~W.~F. Drake and S.~P.~Goldman, Can.~J.~Phys. {\bf 77}, 835 (1999).
\bibitem{plamb} K. Pachucki, J. Phys. B {\bf 35}, 3087 (2002).
\bibitem{Dorrer} C. Dorrer, F. Nez, B. de Beauvoir, L. Julien, and F. Biraben, Phys.
                 Rev. Lett. {\bf 78}, 3658 (1997).
\bibitem{Cancio} P.~C. Pastor, G. Giusfredi, P. De Natale, G. Hagel, C. de
                 Mauro, and M. Inguscio, Phys. Rev. Lett. {\bf 92}, 023001 (2004). 
\bibitem{Lichten} W. Lichten, D. Shiner, and Z.-X. Zhou, Phys. Rev. A {\bf 43}, 1663 (1991).
\bibitem{sans} C.J. Sansonetti and W.C. Martin, Phys. Rev A {\bf 29}, 159 (1984).
\bibitem{nist} P.~J. Mohr and B.~N. Taylor, Rev. Mod. Phys. {\bf 77}, 1 (2005).
\bibitem{fs} K. Pachucki, Phys. Rev. Lett. {\bf 97}, 013002 (2006).
\bibitem{vassen} K.A.H. van Leeuwen and W. Vassen, Europhys. Lett. {\bf 76}, 409 (2006).
\end{thebibliography}
\end{document}